\documentclass[sort&compress,5p]{elsarticle}
\usepackage{graphicx,natbib,amssymb}
\usepackage[pdftex,colorlinks,citecolor=blue,bookmarks]{hyperref}
\usepackage{bm}
\usepackage{amsmath}
\usepackage{color}

\journal{Physics Letter B}

\begin{document}
\begin{frontmatter}
\title{Ground-state properties of even and odd Magnesium isotopes in a symmetry-conserving approach} 

\author[UAM]{Marta Borrajo}

\author[UAM]{J. Luis Egido\corref{cor1}}
\ead{j.luis.egido@uam.es}
\cortext[cor1]{Corresponding author}

\address[UAM]{Departamento de F\'isica Te\'orica, Universidad Aut\'onoma de Madrid, E-28049 Madrid, Spain}

\begin{abstract}
 We present a self-consistent theory for odd nuclei with exact blocking and particle number and angular momentum projection. The demanding treatment of the pairing correlations in a variation-after-projection approach as well as the explicit consideration of the triaxial deformation parameters in a projection after variation method, together with the use of the finite-range density-dependent Gogny force,  provides an excellent tool for the description of odd-even and even-even nuclei.   We apply the theory to the Magnesium isotopic chain and obtain an outstanding description of the ground-state properties,  in particular binding energies, odd-even mass differences, mass radii and electromagnetic moments among others.

\end{abstract}


\end{frontmatter}

In the last years there has been an important development in the description of even-even 
nuclei with effective interactions, in particular with the Skyrme, Gogny and relativistic  \cite{BHR.03,NVR.11,EG.16} ones.
The breakthrough has been possible by means of the beyond-mean-field theories (BMFT), namely by
the recovery of the symmetries broken in the mean-field approach (MFA)  and by the explicit consideration of
large-amplitude fluctuations around the most probable mean-field values. The shape parameters $(\beta,\gamma)$\cite{Skyrme,RE.10,relativistic} (and pairing gaps \cite{Nuria,Nuria2,Nuria3}) were used as coordinates in the framework of the generator-coordinate method (GCM) and the particle-number (PN) and angular-momentum (AM) symmetries were recovered by means of projectors. These developments are called symmetry-conserving configuration mixing  (SCCM) approaches and  have been applied to even-even nuclei. Methods based on the Bohr collective Hamiltonian have also made large progress lately \cite{Gogny_B, Skyrme_B, relativistic_B}.   

Odd nuclei, on the other hand, are far more complicated to deal with.  Even at the mean-field level like in the  Hartree-Fock-Bogoliubov (HFB) or BCS theories,  odd nuclei are numerically  cumbersome and to calculate ground states one must try several spins, parity, etc.  Furthermore,  the blocked structure of the wave function entail the breaking of the time-reversal symmetry and triaxial calculations must be performed. 
The SCCM developments have taken place for even-even nuclei and it seems natural to extend these approaches to odd-even and odd-odd nuclei. As a matter of fact angular-momentum projected calculations  for odd-A nuclei started long ago, though they have been mostly performed on HF or HFB states  in small valence spaces \cite{BGR.65,GW.67,RPK.93,HI.84,HSG.85}. More recently a GCM mixing based on parity and AM-projected  Slater determinants in a model space of antisymmetrized Gaussian wave packets has been carried out in the frameworks of fermionic \cite{NF.08} and antisymmetrized \cite{KTK.13,KK.10}  molecular dynamics. In the latter calculations, however,  the pairing correlations are not treated properly.  A first  extension of BMFT from even to odd nuclei with the Skyrme force has been done recently  in Ref.~\cite{Bally}.  
  
   The purpose of this Letter is to report on the first systematic  description of the odd and even nuclei  of an isotopic chain in a symmetry-conserving approach with the Gogny force in a BMFT considering the $(\beta, \gamma)$ degrees of freedom explicitly and dealing optimally with the pairing correlations. Our approach considers exact triaxial self-consistent blocking and exact particle number and 
   angular momentum conservation. As an illustration of our approach we have chosen the Magnesium isotopic chain for which there is abundant experimental data.
   Basic properties like odd-even mass differences, magnetic and quadrupole moments as well as mass radii, among others, are investigated.

Our starting approach is the HFB theory \cite{RS.80}. As a mean-field approximation  the  HFB  wave function $|\phi\rangle$ is a product of  quasi-particles $\alpha_{\rho}$ defined by  the general Bogoliubov transformation
\begin{equation}  \label{bogtrans}
\alpha^{\dagger} _\rho=\sum_{\mu}U_{\mu \rho}^{}c_{\mu}^{\dagger}+V_{\mu \rho}^{}c_{\mu},
\end{equation}
where ${c_{\mu}^{\dagger},c_{\mu}}$ are the particle-creation and -annihilation operators in the reference  basis, in our case the Harmonic Oscillator one. $U$ and $V$ are the Bogoliubov matrices  to be determined by the Ritz variational principle.   

In our approach we have imposed three discrete self-consistent symmetries on our basis states $\{c_{\mu}^{\dagger},c_{\mu}\}$: spatial parity, $\hat{P}$, simplex,  $\Pi_{1}=\hat{P}e^{-i\pi J_{x}}$ and the  $\Pi_{2} {\cal T}$ symmetry, with $\Pi_{2}= \hat{P}e^{-i\pi J_{y}}$ and  ${\cal T}$ the time reversal operator. 
The first two symmetries provide good parity and simplex quantum numbers and the third allows to use only real quantities. The simplex symmetry furthermore allows to characterize the blocking structure of odd and even nuclei \cite{Ma.75,EMR.80}.  Our basis is symmetrized  in such a way that
\begin{equation}
\Pi_{1} c^{\dagger}_k  \Pi_{1}^{\dagger}  = +i  c^{\dagger}_k,   \;\;\;
\Pi_{1} c^{\dagger}_{\overline k}  \Pi_{1}^{\dagger}  = -i  c^{\dagger}_{\overline k}.
\end{equation}
with ${ k=1,...,M}$ and $2M$ the dimension of the configuration space. We use latin indices  to distinguish the levels according to their simplex, $\{k,l,m\}$ for simplex $+i$ and $\{{\overline k},{\overline l},{\overline  m}\}$ for simplex $-i$. The greek indices 
 ${\mu,\rho}$, on the other hand, do not distinguish simplex and run therefore over the whole configuration space.  
If we further assume that the intrinsic wave function is an eigenstate of the simplex operator, then, for a paired even-even nucleus  half of the quasiparticle operators $\alpha^{\dagger}_{\mu}$, have simplex $+i$ and the other half 
have simplex $-i$, i.e., Eq.~(\ref{bogtrans}) separates in two blocks~:
\begin{eqnarray}  \label{bogtrans_simp} 
\alpha _m^{\dagger} &= &\sum_{k=1}^{M}U_{km}^{+}c_k^{\dagger} +V_{km}^{+}c_{\overline k}^{}, \nonumber \\
\alpha _{\overline m}^{\dagger} &= &\sum_{k=1}^{M}U_{km}^{-}c_{\overline k}^{\dagger} +V_{km}^{-}c_k^{}, 
\end{eqnarray}
with ${ m=1,...,M}$ in an obvious notation. 

The  wave function of the ground state of an even-even nucleus is given by\footnote{The quasiparticle operators that annihilate trivially the particle vacuum are to be omitted from the product.} 
\begin{equation}
| \phi \rangle = \prod_{\mu=1}^{2M} \alpha_\mu  |-\rangle,
\label{Eq:vac}
\end{equation}
with $|-\rangle$ the particle vacuum.  The quasiparticle vacuum $|\phi\rangle$ is obviously defined by
\begin{equation}
\alpha_{\mu}  | \phi \rangle = 0, \;\;   \mu=1,...,2M.
\label{Eq:qp_vac}
\end{equation}
The ground state of an even-even nucleus has simplex $+1$. The quasiparticle excitations 
\begin{equation}
| {\tilde \phi} \rangle =  \alpha^{\dagger}_{\rho_1} |\phi \rangle
\label{Eq:exc}
\end{equation}
correspond to  odd-even nuclei. They can be written as  vacuum
to the quasiparticle operators ${\tilde \alpha}_\rho$, 
\begin{equation}
{\tilde \alpha}^{}_\rho| {\tilde \phi} \rangle =0, \;\;   \rho=1,...,2M.
\label{Eq:vacexc}
\end{equation}
The $2M$ operators  $\{ {\tilde \alpha}^{\dagger}_\rho\}$ are obtained from the set $\{ { \alpha}^{\dagger}_\mu \}$ by replacing the creation operator $\alpha^{\dagger}_{\rho_1}$ by the annihilation operator $\alpha^{}_{\rho_1}$, the other $2M-1$  operators remain unchanged. The simplex of the state $|{\tilde \phi}\rangle$ is given by $\Pi_1| {\tilde \phi} \rangle = i^{n}| {\tilde \phi} \rangle$, where we have introduced the blocking number $n$. It is $n=1$ if $\alpha^{\dagger}_{\rho_1}$ has simplex $+i$ and
$n=-1$ if   $\alpha^{\dagger}_{\rho_1}$ has simplex $-i$. The unblocked wave function $|\phi\rangle$ is vacuum  to $M$ operators with simplex $+i$ and to $M$ with simplex $-i$. The blocked wave function $|{\tilde \phi} \rangle$ is vacuum to  $M_{+}= M-n$ operators ${\tilde \alpha}^{\dagger}_{m}$ with simplex $+i$ and to $M_{-}= M+n$ operators ${\tilde \alpha}^{\dagger}_{\overline m}$ with simplex $-i$.
\begin{eqnarray}  \label{bogtrans_block} 
{\tilde\alpha} _m^{\dagger} &=& \sum_{k=1}^{M}{\tilde U}_{km}^{+}c_k^{\dagger} +{\tilde V}_{km}^{+}c_{\overline k}^{},
\;\;\;  m=1,...,M_{+},\nonumber \\
{\tilde\alpha}_{\overline m}^{\dagger} &=& \sum_{k=1}^{M}{\tilde U}_{km}^{-}c_{\overline k}^{\dagger} +{\tilde V}_{km}^{-}c_k^{},\;\;\;  m=1,...,M_{-}.
\end{eqnarray}
 The matrices $({\tilde U}^{+},{\tilde V}^{+},{\tilde U}^{-},{\tilde V}^{-})$  are rectangular with $M$ rows and $M_{+}$ or $M_{-}$ columns and according to the transformation $\alpha^{\dagger}_{\rho_1} \rightarrow \alpha^{}_{\rho_1}$,  they are obtained, from the $M\times M$ squared matrices  $({U}^{+},{V}^{+},{U}^{-},{V}^{-})$ from Eq.~(\ref{bogtrans_simp}) by the corresponding columns exchange.
    
  Though the state $|{\tilde \phi} \rangle$ has the right blocking structure, since the Bogoliubov transformation mixes creator and annihilator operators and states with different angular momenta, $|{\tilde \phi} \rangle$ is not an eigenstate of the PN   or the AM operators.  As with even-even nuclei, to recover the particle-number symmetry one has to project to the right quantum numbers, see \cite{RS.80}. The {\em easiest} way would be to minimize the HFB energy, i.e., determine $({\tilde U},{\tilde V})$ and then perform the projections, i.e. the so-called projection-after-variation (PAV). The {\em optimal}  way is to determine $({\tilde U},{\tilde V})$ directly from the minimisation  of the projected energy, i.e, the variation-after-projection (VAP) method. From even-even nuclei one knows that PN-VAP is feasible  while AM-VAP is  very
CPU-time consuming.  The approach of solving  the PN-VAP variational equation to find the self-consistent minimum and afterwards to perform an  AM-PAV is not very good because the AMP is not able to exploit any degree of freedom of the HFB transformation and self-consistency with respect to the AMP is not guarantied. An intermediate way is to perform an approximate AM-VAP approach by solving the variational PN-VAP equation for a large set of relevant physical situations as to cover the sensitive degrees of freedom.  Afterwards an AM-PAV to this set of wave functions will  determine the absolute minimum among these states for different angular momenta. Usually it is believed that the strongest energy dependence of the nuclear interaction is related to the deformation parameters $(\beta,\gamma)$ and we will consider them as the additional degrees of freedom.  Notice that this method guarantees, at least, AM-VAP self-consistency with respect to these relevant  quantities. Therefore, in order to obtain a grid of wave functions we solve the PN-VAP  constrained equations
\begin{eqnarray}
{E^{\prime}}[{\tilde \phi}]= \frac{ \langle{\tilde \phi}^{}|\hat{H}\hat{P}^{N}|{\tilde \phi}{} \rangle}{\langle{\tilde \phi}^{}|\hat{P}^{N}|{\tilde \phi}^{} \rangle} -  \langle {\tilde \phi} |\lambda_{q_{0}}\hat{Q}_{20} + \lambda_{q_{2}}  \hat{Q}_{22} | {\tilde \phi} \rangle, \label{E_Lagr_bet-gam}
\end{eqnarray}
with  the Lagrange multiplier $\lambda_{q_{0}}$  and $\lambda_{q_{2}}$ being determined by the constraints 
\begin{equation}
\langle {\tilde \phi} |\hat{Q}_{20} | {\tilde \phi} \rangle =q_{0}, \;\; \; \langle {\tilde \phi} |\hat{Q}_{22} | {\tilde \phi} \rangle =q_{2}. \label{q0_q2_constr}
\end{equation}
The relation between $(\beta,\gamma)$ and $(q_{0},q_{2})$ is given by 
$\beta= \sqrt{20\pi(q_{0}^{2} + 2 q_{2}^{2})}/3r^{2}_{0}A^{5/3}$,
$\gamma = \arctan (\sqrt{2}{q_2}/q_{0})$
with $r_{0}=1.2$ fm and  $A$ the mass number. 

In this work we are interested in the odd-even Magnesium isotopes. We therefore consider  wave functions of the form
\begin{equation} \label{eq:odd_even_ansatz}
|{\tilde \phi}^{\pi} \rangle = \alpha_{\rho_{1}}^{\dagger}\prod_{\mu=1}^{2M} \alpha_{\mu}|-\rangle.
\end{equation} 
According to the isospin and parity we have four blocking channels: protons (neutrons) of positive or negative parity. Since Magnesium isotopes have $Z=12$, we restrict ourselves to the neutron channels. Notice that in the running product of Eq.~(\ref{eq:odd_even_ansatz}), orbitals with the same parity are occupied pairwise, therefore the parity, $\pi$, of the state  $|{\tilde \phi} \rangle$ is given by the parity of the blocked level $\alpha^{\dagger}_{\rho_{1}}$.
One can furthermore block a state with positive or negative simplex, but since we do not break time reversal explicitly both possibilities are degenerated.

The minimization of Eqs.~(\ref{E_Lagr_bet-gam}-\ref{q0_q2_constr}) is performed with the conjugated-gradient method \cite{grad}. The blocking structure of the wave function of Eq.~(\ref{eq:odd_even_ansatz}) is a self-consistent symmetry and for a given blocking number we determine the lowest solution in the blocked channel compatible with the imposed constraints.  That is,  it does not matter which level is initially blocked, at the end of the iteration process the PN-VAP energy and the HFB wave function are independent of this election.
\label{Sect:PES}

The next step is the simultaneous particle-number and angular-momentum projection (PNAMP) of each state   $|{\tilde\phi}^{\pi}(\beta,\gamma)\rangle$  that conforms the $(\beta,\gamma)$ grid, 
\begin{eqnarray}
|\Psi^{N,I,\pi}_{M,\sigma } (\beta,\gamma) \rangle &=&   \sum_{K} g^{I}_{K\sigma} P^N P^I_{MK} \; |{\tilde\phi}^{\pi} (\beta,\gamma)\rangle  \nonumber \\
& =&   \sum_{K} g^{I}_{K\sigma}  |IMK,\pi,N, (\beta,\gamma) \rangle,  \label{eq:GCM_Ansatz_bet_gam}
\end{eqnarray}
where the  coefficients  $g^{I}_{K\sigma}$  are variational parameters.   They are determined by the energy minimization which  provides a reduced Hill-Wheeler-Griffin \cite{HWG} equation 
\begin{equation}
\sum_{K'} \, \,(\mathcal{H}^{N,I,\pi}_{K, K'} - E^{N,I,\pi}_\sigma \mathcal{N}^{N,I,\pi}_{K,K'}) g^{I}_{K^{\prime}\sigma} = 0.
\label{HW_K}
\end{equation} 
where $\mathcal{H}^{N,I,\pi}_{K K'}$ and $\mathcal{N}^{N,I,\pi}_{K,K'}$ 
are the Hamiltonian and norm overlaps defined by 
\begin{eqnarray}
\mathcal{H}^{N,I,\pi}_{K, K'} \! & = & \!   \langle IMK,\pi,N, (\beta,\gamma) |H | IMK',\pi,N,(\beta,\gamma) \rangle \label{hamove} \\
\mathcal{N}^{N,I,\pi}_{K,K'} \! & = & \!   \langle IMK,\pi,N,(\beta,\gamma) | IMK',\pi,N,(\beta,\gamma) \rangle \label{normove}.
\end{eqnarray}
The presence of the norm matrix in Eq.~(\ref{HW_K}) is due to the non-orthogonality  of the states $|IMK,\pi,N,(\beta,\gamma) \rangle$. Eq.~(\ref{HW_K}) is solved by standard techniques \cite{RS.80}.
Notice that at each $(\beta,\gamma)$ point one can have several eigenvalues $E^{N,I,\pi}_\sigma$ labeled by $\sigma$, $\sigma=0$ corresponds to the lowest solution. 

The solution of Eqs.~(\ref{HW_K}) in the $(\beta,\gamma)$ grid for different angular momenta and parity provides $E^{N,I,\pi}_{\sigma}(\beta,\gamma)$ as a function of $(\beta,\gamma)$, $I$, $\pi$ and $\sigma$. This energy  can be written as
\begin{equation}
E^{N,I,\pi}_\sigma (\beta,\gamma)=\frac{\langle \Psi^{N,I,\pi}_{M,\sigma } (\beta,\gamma)| H  |\Psi^{N,I,\pi}_{M,\sigma } (\beta,\gamma) \rangle }{\langle \Psi^{N,I,\pi}_{M,\sigma } (\beta,\gamma)| \Psi^{N,I,\pi}_{M,\sigma } (\beta,\gamma) \rangle },
\label{Eq:PESIpi}  
\end{equation}
which obviously  represents the potential energy surface (PES) of the projected energy in the $(\beta,\gamma)$ plane     
for the given quantum numbers. This projected PES differs from the usual mean field PES and are angular momentum $(I)$, parity $(\pi)$ and state $(\sigma)$ dependent.  The minimum value of $E^{N,I,\pi}_{\sigma}(\beta,\gamma)$ in the PES provides the energy and  the deformation parameters $(\beta_{\rm min},\gamma_{\rm min})$ of the state characterized by the quantum numbers $(I,\pi,\sigma)$ in this approximation. Its wave function is given by  $|\Psi^{N,I,\pi}_{M,\sigma } (\beta_{\rm min},\gamma_{\rm min}) \rangle$. 

Since the states $|IMK,\pi,N,(\beta,\gamma) \rangle$ are not orthogonal, the weights  $g^{I}_{K\sigma}$
do not satisfy  $\sum_{K} |g^{I}_{K\sigma}|^2=1$. The collective wave function
\begin{equation}
G^{I}_{K,\sigma} = \sum_{K'} \large( \mathcal{N}^{N,I,\pi}\large)_{K,K'}^{1/2} g^{I}_{K',\sigma}, 
\label{G_KK}
\end{equation}
on the other hand, does and can be interpreted as a probability amplitude.

\begin{figure*}[t]
	\begin{center}
		\includegraphics[angle=0,scale=.9]{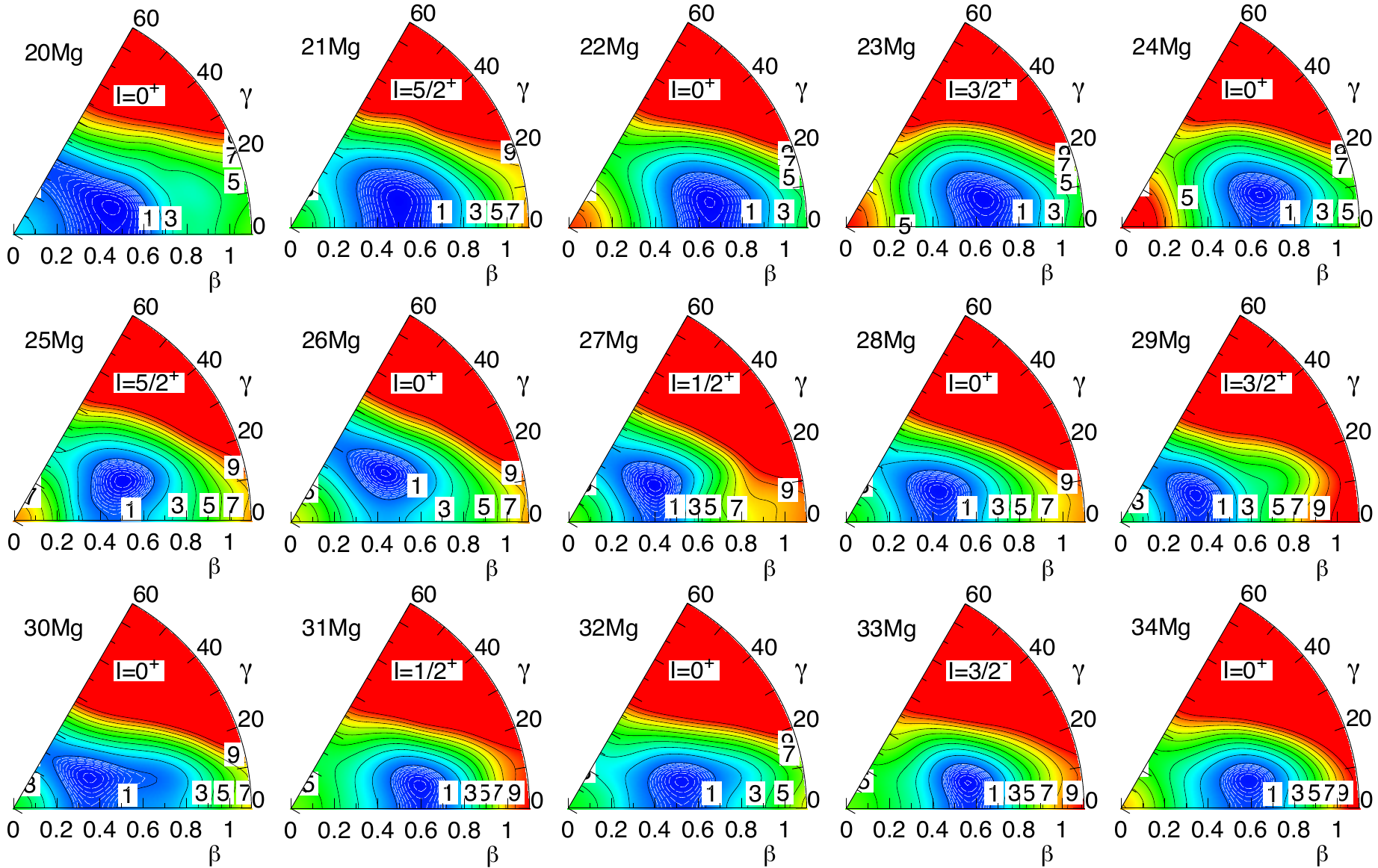}
		\caption{Contour plots of $E^{N,I,\pi}_{\sigma=0}(\beta,\gamma)$, see Eq.~(\ref{HW_K}), as a function of $(\beta,\gamma)$ for positive parity and for the angular momentum $I$ providing the lowest energy. The  solid black contour lines start at 1MeV and increase 1 MeV. The dashed white lines start at zero and increase  0.1 MeV. The zero contour is only present if the minimum is flat enough. The angle $\gamma$ units are degrees.}
		\label{PES_POS}       
	\end{center}
\end{figure*}

 In the calculations the intrinsic many body wave functions $|{\tilde \phi}(\beta,\gamma)\rangle$ are expanded in a Cartesian harmonic oscillator basis and the number of spherical shells included in this basis is $N_{shells}=8$ with an oscillator length of $b=1.01A^{1/6}$. The $(\beta,\gamma)$ grid of equilateral triangles contains 116 points. The  angular momentum projection has been done with the set of integration points in the Euler angles  $(N_{\boldsymbol{\alpha}}=N_{\boldsymbol{\beta}}=N_{\boldsymbol{\gamma}}=32)$ in the intervals $\boldsymbol{ \alpha} \in [0,2\pi ], \boldsymbol{ \beta} \in [0,\pi],\boldsymbol{\gamma} \in [0,2\pi]$. The number of points to perform the integral of the particle-number projection is  $11$. 
In the calculations we use the Gogny interaction \cite{DG.80} with the D1S parameterization \cite{BGG.91}.
We consider all exchange terms of the interaction, the Coulomb force and the two-body correction of the kinetic energy to avoid problems with the PNP \cite{AER.01Ex}. Concerning the density dependence of the force we adopt  the projected density prescription for the PNP and the mixed one for the AMP, for further details see for example Ref.\cite{RE.10,EG.16}.  

To illustrate the method we have applied the discussed theory to the calculation of the  bulk properties of the Magnesium isotopes. Towards this end we have to determine the wave function of the ground state of each isotope. This is done in the following way. 

{\em Step} 0: We choose a parity (positive for example) for the blocked state in Eq.~(\ref{eq:odd_even_ansatz}). Next we solve the PN-VAP variational equations Eqs.~(\ref{E_Lagr_bet-gam}, \ref{q0_q2_constr}) for all 
$(\beta,\gamma)$ values of the grid. This step provides a set of wave functions $|{\tilde \phi}^{\pi}(\beta,\gamma)\rangle$  ($P^N|{\tilde \phi}^{\pi}(\beta,\gamma)\rangle$)  with the right   parity (and particle number). However, they are not eigenstates of the angular-momentum operator.

{\em Step} 1.0:  We choose a value for the angular momentum, $1/2$ for example.  We now solve Eq.~(\ref{HW_K}) for all  $|{\tilde \phi}^{\pi}(\beta,\gamma)\rangle$ of the grid determined in step 0 for the given $I$-value.  This provides the PES of Eq.~(\ref{Eq:PESIpi}). 
The minimum  value of $E^{N,1/2,+}_{\sigma=0}(\beta,\gamma)$ provides the  $(\beta^{1/2,+}_{\rm min},\gamma^{1/2,+}_{\rm min})$ values. 

{\em Step} 1.1:  We repeat step 1.0 for all $I$-values,  and determine the corresponding PESs and 
the $(\beta^{I,\pi}_{\rm min},\gamma^{I,\pi}_{\rm min})$ values for $I= 3/2,5/2, ...$ .  When this step is completed we have found the minima $(\beta^{I,\pi}_{\rm min},\gamma^{I,\pi}_{\rm min})$ for $I=1/2, 3/2, 5/2, ...$  and positive parity.  Their corresponding energies are  $E^{N,1/2,+}_{\sigma=0}(\beta^{1/2,+}_{\rm min},\gamma^{1/2,+}_{\rm min})$, $E^{N,3/2,+}_{\sigma=0}(\beta^{3/2,+}_{\rm min},\gamma^{3/2,+}_{\rm min})$, etc.  From this set of energies
the smallest one provides the angular momentum of the lowest state with positive parity, which we
call $I_1$, and its energy $E^{N,I_1,+}_{\sigma=0}(\beta^{I_1,+}_{\rm min},\gamma^{I_1,+}_{\rm min})$.

{\em Step} 2:  We repeat steps 0,  and 1  for the other parity (negative). When this step is completed we have determined the corresponding PESs,  the deformation parameters of the minima and
the  energies $E^{N,1/2,-}_{\sigma=0}(\beta^{1/2,-}_{\rm min},\gamma^{1/2,-}_{\rm min})$, $E^{N,3/2,-}_{\sigma=0}(\beta^{3/2,-}_{\rm min},\gamma^{3/2,-}_{\rm min})$, etc.  As before the smallest energy  provides the angular momentum of the lowest state with negative parity. We call it $I_2$ and its energy $E^{N,I_2,-}_{\sigma=0}(\beta^{I_2,-}_{\rm min},\gamma^{I_2,-}_{\rm min})$.

 The smallest value of $E^{N,I_1,+}_{\sigma=0}(\beta^{I_1,+}_{\rm min},\gamma^{I_1,+}_{\rm min})$ and  $E^{N,I_2,-}_{\sigma=0}(\beta^{I_2,-}_{\rm min},\gamma^{I_2,-}_{\rm min})$ provides the binding energy, the spin and the parity of the ground state of the given nucleus as well as  the deformation parameters $(\beta^{I,\pi}_{\rm min},\gamma^{I,\pi}_{\rm min})$.  
  The wave function  $|\Psi^{N,I,\pi}_{M,\sigma=0 } (\beta^{I,\pi}_{\rm min},\gamma^{I,\pi}_{\rm min}) \rangle$ 
 characterized by these quantum numbers determines the wave function of the ground state which will be used to calculate electromagnetic properties, radii and so on.

 Before considering the ground-state properties let us discuss the PESs of the different isotopes since they allow to determine the quality of the approach and  in particular if the energy minimum is well defined.

\begin{figure}[t]
	\begin{center}
		\includegraphics[angle=0,scale=.4]{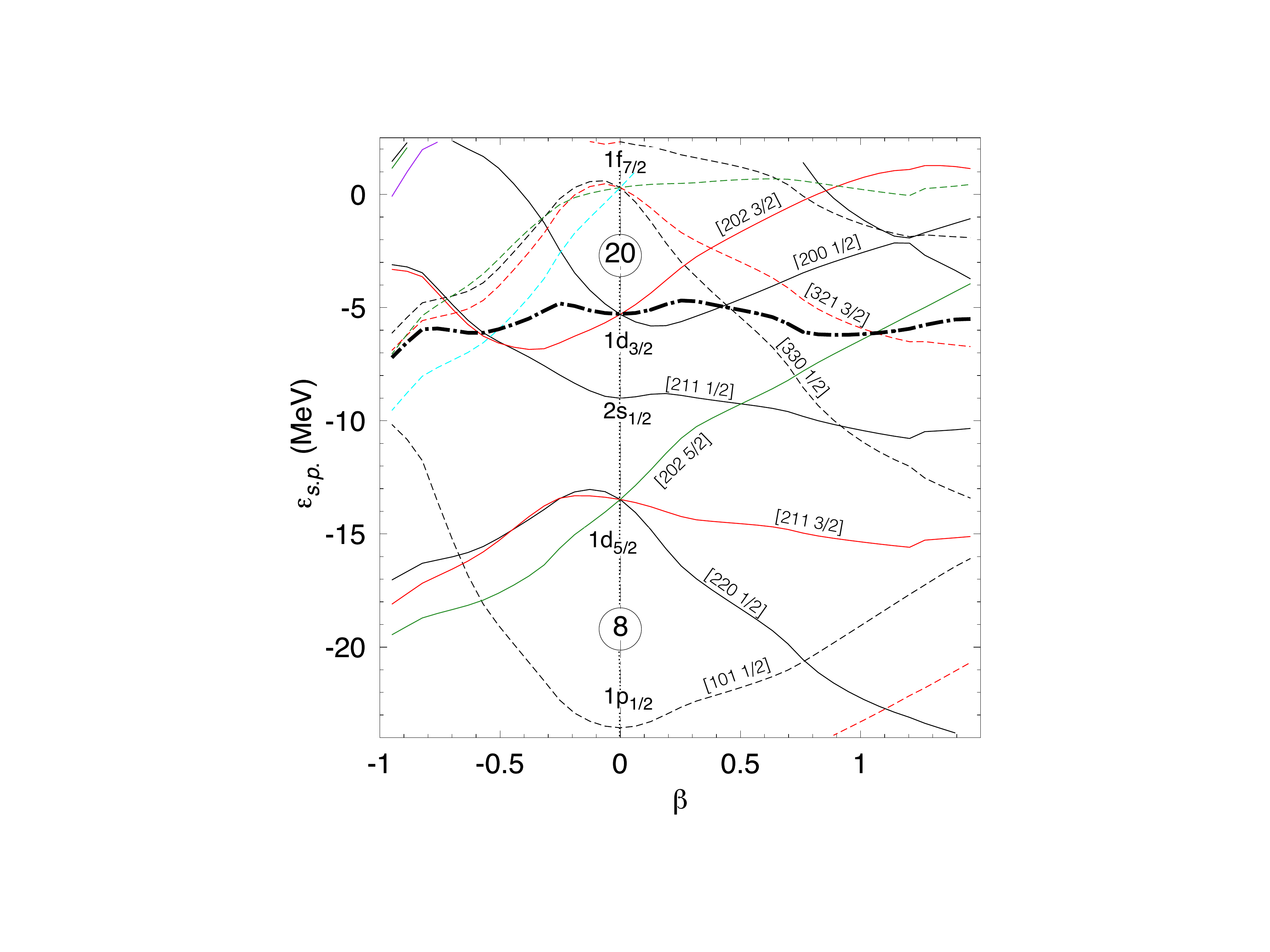}	
		\caption{Single-particle levels of $^{30}$Mg for neutrons obtained from the solution of the axially-symmetric HFB equation. The thick dashed lines represent the corresponding Fermi level.}
		\label{Fig:spe}       
	\end{center}
\end{figure}

In Fig.~\ref{PES_POS}  we present contour lines of the PES $E^{N,I,\pi}_{\sigma=0}(\beta,\gamma)$  in the $(\beta,\gamma)$ plane for the $I^{\pi}$ of the ground state for the Mg isotopes.   Let us first mention that the predicted spins and parities coincide with the experimental values in all cases. Interestingly all nuclei are  triaxial with $\gamma$ values ranging from 10$^{\circ}$ to 25$^{\circ}$, and have large $\beta$ deformations. Since most minima are very well defined we can conclude that our approach of keeping only one point of the $(\beta,\gamma)$ plane works very well for  most nuclei.  The softest  nucleus is  $^{20}$Mg where a GCM in the  $(\beta,\gamma)$  could be performed, which, in general, would lead to smaller deformation than the one quoted here.
In Table~\ref{Table1} the $(\beta,\gamma)$ values of the ground states are listed. For a better understanding of our results we use  the  collective wave function, Eq.~(\ref{G_KK}),  to obtain the  $|K|$ distribution of the odd neutron.  The  $|K|$ component with the largest weight is listed in Table~\ref{Table1}  and it turns out that these wave functions have rather pure $|K|$. This purity, in spite of the, sometimes, large triaxiality has been also observed for even-even nuclei (see Table I of Ref.~\cite{RE.10}) for the ground state band. The  absence of K-mixing is probably due to the low level density of light nuclei.    Furthermore we analyze the intrinsic HFB wave function $|{\tilde \phi}^{\pi}(\beta,\gamma)\rangle$ in the canonical basis what provides information on the quantum numbers of the blocked state for odd systems.  To  guide the discussion we will use a Nilsson plot, see Fig.~\ref{Fig:spe} for the particular case of $^{30}$Mg.  We will furthermore use in our analysis  the particle plus rotor  (PR) model.   Let us first discuss the spin values and parities.
 In the PR model, and according to the deformations of the Mg isotopes,  one expects to be in the strong-coupling limit (strong deformations), in which case the lowest possible spin is $I_{\sigma=0}= K$, or in the decoupling limit (intermediate deformations), in which case $I_{\sigma=0}= j$.  According to Table~\ref{Table1} and Fig.~\ref{Fig:spe}, the nucleus  $^{21}$Mg has a very pure $|K|=1/2$ character and consequently a large component of the wave function of the last neutron is in the orbital $[220 \;1/2]$ of the 1d$_{\frac{5}{2}}$ subshell.  The theoretical value for the spin and parity of $^{21}$Mg  is $I^{\pi}=\frac{5}{2}^{+}$ which agrees with the decoupling limit prediction of $I=j= \frac{5}{2}$ and with the experimental data. This is a bit surprising since the
 $\beta$ value is rather large and in principle one would expect the strong coupling limit. A look at the experimental data reveals that the $I=|K|= \frac{1}{2}$ state is just 200 keV above the $I=j= \frac{5}{2}$ one. As a matter of fact  the
 $^{23,25 }$Mg  isotopes  with $|K|=3/2$ ($[211\; 3/2]$ orbital) and  $|K|=5/2$  ($[202\; 5/2]$ orbital), with a larger deformation, see Table~\ref{Table1},  do have  $I=K= \frac{3}{2}$  and  $I=|K|= \frac{5}{2}$, respectively, in agreement with the experimental values.  The nucleus  $^{27}$Mg with a neutron with $|K|= \frac{1}{2}$ in the
  2s$_{1/2}$ sub-shell has obviously $I=\frac{1}{2}$ in agreement with the experimental value.   In the case of  $^{29}$Mg we have $|K|=1/2$ and  the odd neutron sits in the orbital $[200 \;1/2]$.  Since its  deformation is $\beta= 0.37$, smaller than the one of  $^{21}$Mg, we expect also in this case the decoupling limit value of $I= \frac{3}{2}$, in agreement with our result and the experimental data. In the case of $^{31}$Mg,  with  $|K|=1/2$, we have two particles in the $[330 \;1/2]$ and one particle in the $[200 \;1/2]$, see below, as in $^{29}$Mg. However,  in this nucleus the deformation is $\beta= 0.60$.  We are in the strong-coupling limit,  and expect therefore  $I=K= \frac{1}{2}$ in coincidence with the theoretical and the experimental values. All these nuclei have the unpaired nucleon in the 2s or the 1d shells and have positive parity.  Our last odd nucleus, $^{33}$Mg, has  $|K|=3/2$,
 the last neutron sits in the $[321\; 3/2]$ orbital and it has a large deformation. We expect therefore $I=|K|= \frac{3}{2}$ 
and negative parity, in agreement with the theoretical and the experimental values.

\begin{table}
	\centering
	\begin{tabular}{|c|c |c| c| c| c| c|}
		\hline
		$A $  &  $\footnotesize I^{\pi} $  & $\beta,\gamma$ &$\beta_{exp}$ &  $ |K| (\%)$  &  $Q_{spec}$  
		\tabularnewline
		\hline
		$20$ & 	$0^{+}$   & $0.46,17.5^{\circ}$  & ---&--   & ---
		\tabularnewline
		$21$ & 	$\frac{5}{2}^{+}$   & $0.54,14.9^{\circ}$ &---&  $\frac{1}{2} (99.1 \%)$  &   $-17.80$ 
		\tabularnewline
		$22$ & 	$0^{+}$  &     $0.65,12.2^{\circ}$ & 0.58 (11) &   --& ---
		\tabularnewline
		$23$ & 	$\frac{3}{2}^{+}$    & $0.64,10.9^{\circ}$ & ---&  $\frac{3}{2} (99.9 \%)$ &     $13.89$ 
		\tabularnewline
		$24$ & 	$0^{+}$ & $0.65,12.2^{\circ}$ & 0.605 (8) &  --&   ---
		\tabularnewline
		$25$ & 	$\frac{5}{2}^{+}$   & $0.54,17.5^{\circ}$ &---&  $\frac{5}{2} (99.7 \%)$  &    $22.47$ 
		\tabularnewline
		$26$ & 	$0^{+}$  & $0.49,25.3^{\circ}$ & 0.482 (10)& --&  ---
		\tabularnewline 
		$27$ & 	$\frac{1}{2}^{+}$   & $0.41,23.4^{\circ}$ & -&  $\frac{1}{2} (100 \%)$ &   0 
		\tabularnewline 
		$28$ & 	$0^{+}$   & $0.46,17.5^{\circ}$ & 0.491 (35)& --& ---
		\tabularnewline
		$29$ & 	$\frac{3}{2}^{+}$  & $0.37,19.1^{\circ}$ & --- &  $\frac{1}{2} (96.0 \%)$ &    $-10.71$ 
		\tabularnewline 
		$30$ & 	$0^{+}$  & $0.39,21.1^{\circ}$ & 0.431 (19)& --& ---
		\tabularnewline 
		$31$ & 	$\frac{1}{2}^{+}$  & $0.60,11.7^{\circ}$ & ---&  $\frac{1}{2} (100.0 \%)$ & 0
		\tabularnewline
		$32$ & 	$0^{+}$   & $0.54,14.9^{\circ}$ & 0.473(43)&--& ---
		\tabularnewline
		$33$ & $\frac{3}{2}^{-}$&   $0.60,11.7^{\circ}$&--- &   $\frac{3}{2} (99.9 \%)$   &  $14.17$
		\tabularnewline
		$34$ & 	$0^{+}$    & $0.62,13.0^{\circ}$ &0.58(6)& --&---
		\tabularnewline		
		\hline  
	\end{tabular}
	\caption{The 2nd and 3rd columns display the spin and parity and the $\beta,\gamma$ deformations of the ground state of the different isotopes. Notice that only $^{33}$Mg has a ground state with negative parity. The 4th column shows the experimental $\beta$ deformation taken from Refs.~\cite{Ra.01,Iwa.01}. The 5th  column lists the $|K|$ component with the largest weight in the wave function, see Eq.~(\ref{G_KK}), with the percentage of this $|K|$ value in the total wave function. The  6th
column provides	the  theoretical spectroscopic quadrupole moments, in efm$^{2}$. }
	\label{Table1}
\end{table}

We now discuss the shapes of the nuclei. The nucleus  $^{20}$Mg has a neutron shell closure at $N=8$ and therefore one expects a smaller 
deformation than for the heavier isotopes. The same behaviour is expected for  $^{21}$Mg with just one neutron outside the closed shell. The  isotopes $^{22-24}$Mg have a $\beta$-value close to $0.65$ and correspond to the filling of the Nilsson orbitals $[220\;1/2]$ and $[211\;3/2]$ of the  d$_{5/2}$ sub-shell,
see Fig.~\ref{Fig:spe}, which are down-sloping.  The  orbital $[202\;5/2]$ of the 
 d$_{5/2}$ sub-shell starts being occupied in $^{25}$Mg which causes a decrease of the deformation because of its up-sloping character. 
 \begin{figure*}[t]
 	\begin{center}
 		\includegraphics[angle=0,scale=.55]{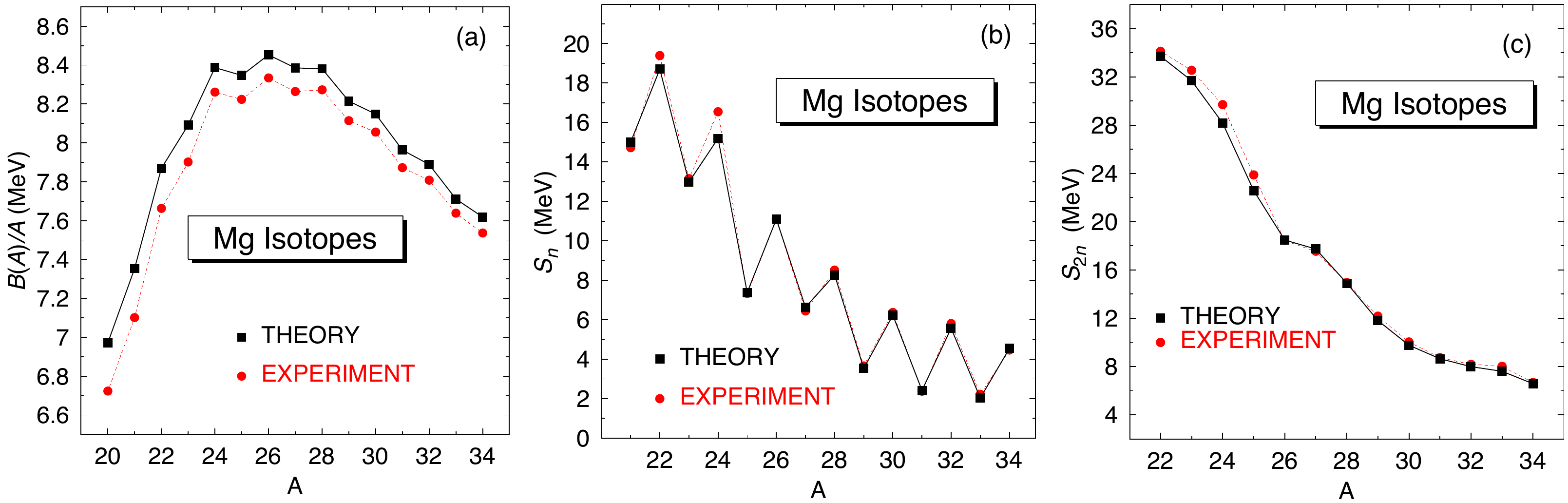}
 		\caption{  (a) Binding energy per particle versus de mass number.  (b) One-neutron separation energies versus  the mass number. (c) Two-neutron separation energies versus the mass number. The experimental values are taken from Ref.~\cite{Brook_data}}.
 		\label{fig:Ener_con}       
 	\end{center}
 \end{figure*}

 The nuclei $^{26-28}$Mg correspond to the filling up of the d$_{5/2}$  and  s$_{1/2}$
sub-shells and the calculated  $\beta$-value is $0.45$ which is close to the crossing of the $[202\;5/2]$ and the $[211\;1/2]$ Nilsson levels. If we now add more neutrons we populate the orbital $[200\; 1/2]$ of the d$_{3/2}$ sub-shell which is down-sloping for small and up-sloping for larger $\beta$-values. This explains the moderate deformation of $^{29-30}$Mg.  The nearest orbitals available to host the next neutrons are the up-sloping  $[202\;3/2]$ of the d$_{3/2}$ sub-shell and the strongly down-sloping   $[330\; 1/2]$ of the f$_{7/2}$ sub-shell.  In this case it is  energetically most convenient  to start filling the  $[202\;3/2]$ orbital at moderate deformation.  It should be noticed, however,  the softness of the PES of $^{30}$Mg in the $\beta$ degree of freedom corresponding to the population of the $[330\; 1/2]$ orbital  at larger deformation. In the PES of  $^{31}$Mg we observe  an abrupt increment of the  deformation parameter as compared with $^{30}$Mg. This is because now the orbital  $[330\; 1/2]$ is filled and in the orbital $[202\;3/2]$ there is only one neutron,  indicating the beginning of the inversion island \cite{ME.16}. For heavier isotopes the up-sloping character of the  $[202\;3/2]$ orbital at larger deformations will favour the filling of the $[321\; 3/2]$ orbital of the f$_{7/2}$ shell, driving these isotopes to even larger deformations as we obtain for the $^{32-34}$Mg isotopes. 

As mentioned, all analysed Mg  isotopes are triaxial and, with the exception of $^{26}$Mg, 
  rather soft towards the prolate axis, i.e., contour lines less than 1 MeV  cross the prolate axis. These nuclei, because of their large $\beta$ values, are much harder towards  oblate shapes. The softest ones  are those with the smallest deformation parameter $\beta$, namely $^{26,27}$Mg  and $^{29,30}$Mg for which the contour lines less than 2 MeV  cross the oblate axis. Furthermore, the experimental deformations listed in Table~\ref{Table1} are in good agreement with the theoretical values. Notice, however, that at variance with our values, the experimental deformations have been extracted from $E2$ transition probabilities, see Refs.~\cite{Ra.01,Iwa.01}.

We now discuss relevant properties of the ground states.
In panel (a) of Fig.~\ref{fig:Ener_con} we present the theoretical binding energies  per particle for the Mg isotopes together with the experimental ones versus the mass number. The theoretical binding energies have been obtained from the energy minima of the corresponding ground state PESs. 
The theory line follows very closely the general behaviour of the experimental one.  We obtain overbinding which is due to the fact that we are using the D1S parameterisation of the Gogny force which was fitted to reproduce experimental data with the HFB method.  Though the authors of Ref.~\cite{DG.80} left some room  for eventual BMF effects  apparently this was not sufficient, see also Refs.~\cite{AMEDEE,Masses_tomas}. One should furthermore consider that the 8 harmonic oscillator shells used in the calculations are alright  to provide relative  but not  absolute energies for which  a larger number of shells is needed, see Ref.~ \cite{Masses_tomas,Ray_Mg_04}. Based on these references one can estimate that an additional overbinding of 2.3 to 2.7 MeV should be added to the results of the present calculations.

 In this plot one can appreciate the odd-even staggering in the two parabolas,
one for even-even and another for the odd-even isotopes, obtained both in the experiment and in the theory. The parabola maximum  at $A=26$ corresponds to the neutron half-shell, $N=14$, which provides maximal binding per particle.   In panels (b) and (c) we present the one-  and two-neutron separation energies, respectively.
For $S_n$, with the exception of  two isotopes, $^{22,24}$Mg, we obtain an extraordinary agreement between the theoretical results and the experimental data. The small disagreement observed for the nuclei $^{22,24}$Mg is probably related to the fact that proton-neutron pairing is not included in our calculations.  Therefore, we find the largest discrepancy in $^{24}$Mg
corresponding to the $N=Z=12$ case.  For $^{22}$Mg the disagreement is smaller and for $^{26}$Mg, with the neutron 1d$_{5/2}$  subshell closure, the p-n pairing looses relevance.
In the $S_{2n}$ case the excellent agreement is maintained but now with the exception of the isotopes $^{22-25}$Mg for which the agreement is not as good as for the others.
The small plateau found at $A=26,27$ is due to the behaviour observed at the top of the parabola in panel (a).

\begin{figure}[t]
	\begin{center}
		\includegraphics[angle=0,scale=.5]{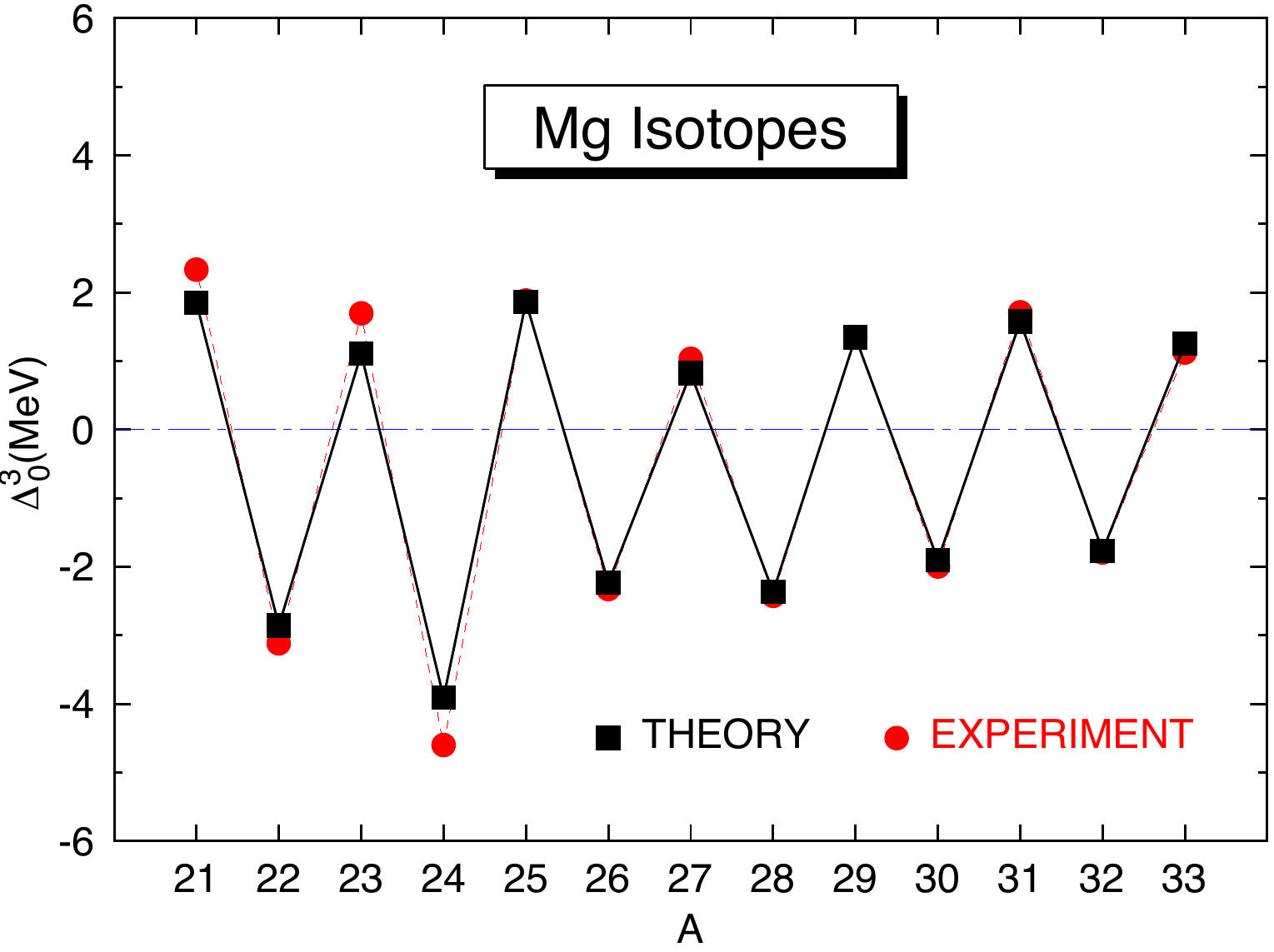}	
		\caption{Odd-even mass differences according to Eq.~(\ref{Eq:oemd}). The experimental data are from Ref.~\cite{Brook_data}}.
		\label{fig:oemd}       
	\end{center}
\end{figure}

In our approach the pairing correlations are treated specially well.  First, the finite range density dependent Gogny force used in the calculations is considered to be one of the best to describe pairing correlations and used as benchmark in many calculations. Second, the use of the PN-VAP approach avoids the pairing collapse in the weak pairing regime which is normally observed in the case of odd-even nuclei. And third, the Coulomb anti-pairing effect (CAP) is taken into account  since all exchange terms of the force, in particular the  Coulomb ones,  are considered in our calculations. A  quantity which allows to extract information  on the pairing energies from the experimental nuclear mass is the odd-even mass difference.  In the three point approach this magnitude is given by
\begin{equation}
\Delta^{3}_{0} (A) = \frac{1}{2}\left[  B(A+1) +B(A-1) -2 B(A)\right],
\label{Eq:oemd}
\end{equation}
with the proton number $Z$ a constant even number and $B(A)$ a  positive number. In Fig.~\ref{fig:oemd} we plot $\Delta^{3}_{0} (A)$ for the Mg isotopes as a function of the mass number. The points above the horizontal line correspond to the odd-even nuclei and those below to the even-even ones.  On average the odd-even nuclei have about 0.5 MeV less pairing  than the even-even ones. The agreement between the theoretical results  and the experimental  data  is excellent, specially for the heavier isotopes. For the lighter 
nuclei, in particular $^{21}$Mg and  $^{23-24}$Mg, the theoretical results are a bit smaller, in absolute value, than the experimental ones. This is again a consequence of the mentioned absence of p-n pairing in our calculations.

\begin{figure}[t]
	\begin{center}
		\includegraphics[angle=0,scale=.55]{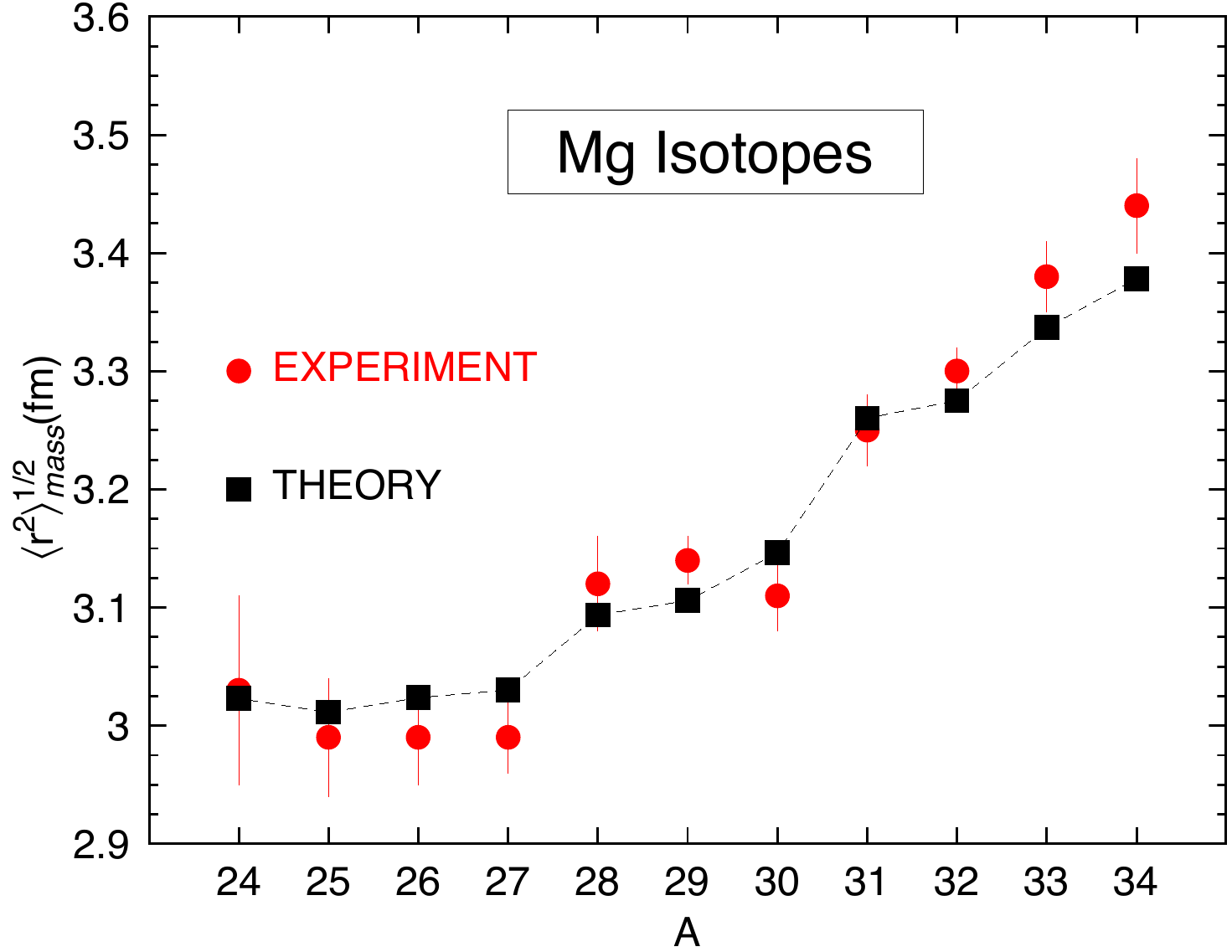}	
		\caption{Radii of the nuclei $^{27-28}$Mg in the PNVAP+PNAMP approach. The experimental data are from Ref.\cite{mass_radii}.}
		\label{fig:radii}       
	\end{center}
\end{figure}

Another relevant quantity is the nuclear radius.   In Fig.~\ref{fig:radii} the experimental  mass radii \cite{mass_radii} corresponding to point mass nucleons\footnote{Private communication of Dr. Shin Watanabe} are plotted
together with the theoretical results.
In the calculation of the mass radius we consider the one-body term of the center-of-mass correction.
The theoretical results reproduce very well the overall experimental behaviour. One can distinguish three well
differentiated regions.  We first observe a rather flat behaviour of the mass radius for $^{24-26}$Mg in which the increase of the neutron radius with filling the neutron
1d$_{\frac{5}{2}}$ orbital is compensated by a compression of the charge distribution. This effect has been observed in the Ne \cite{Ge.08,Ma.11} and in the Mg isotopes \cite{Yor.12}.  Though with the filling of the 2s$_{\frac{1}{2}}$ orbital one would expect an increase of the mass radius, it seems that the mentioned compensation persists also for $^{27}$Mg.    The second region corresponds to the nuclei $^{28-30}$Mg, where we observe a clear increase of the mass radius associated with two neutrons in the 2s$_{\frac{1}{2}}$ or 1d$_{\frac{3}{2}}$ orbitals. The third region, for $A\ge 31$ is marked by the beginning of the inversion island in $^{31}$Mg \cite{ME.16} and the rise in the
mass radius observed for $A\ge 31$  is associated with the increasing  occupation of the 1f$_{\frac{7}{2}}$ orbital.

\begin{figure}[t]
	\begin{center}
		\includegraphics[angle=0,scale=.5]{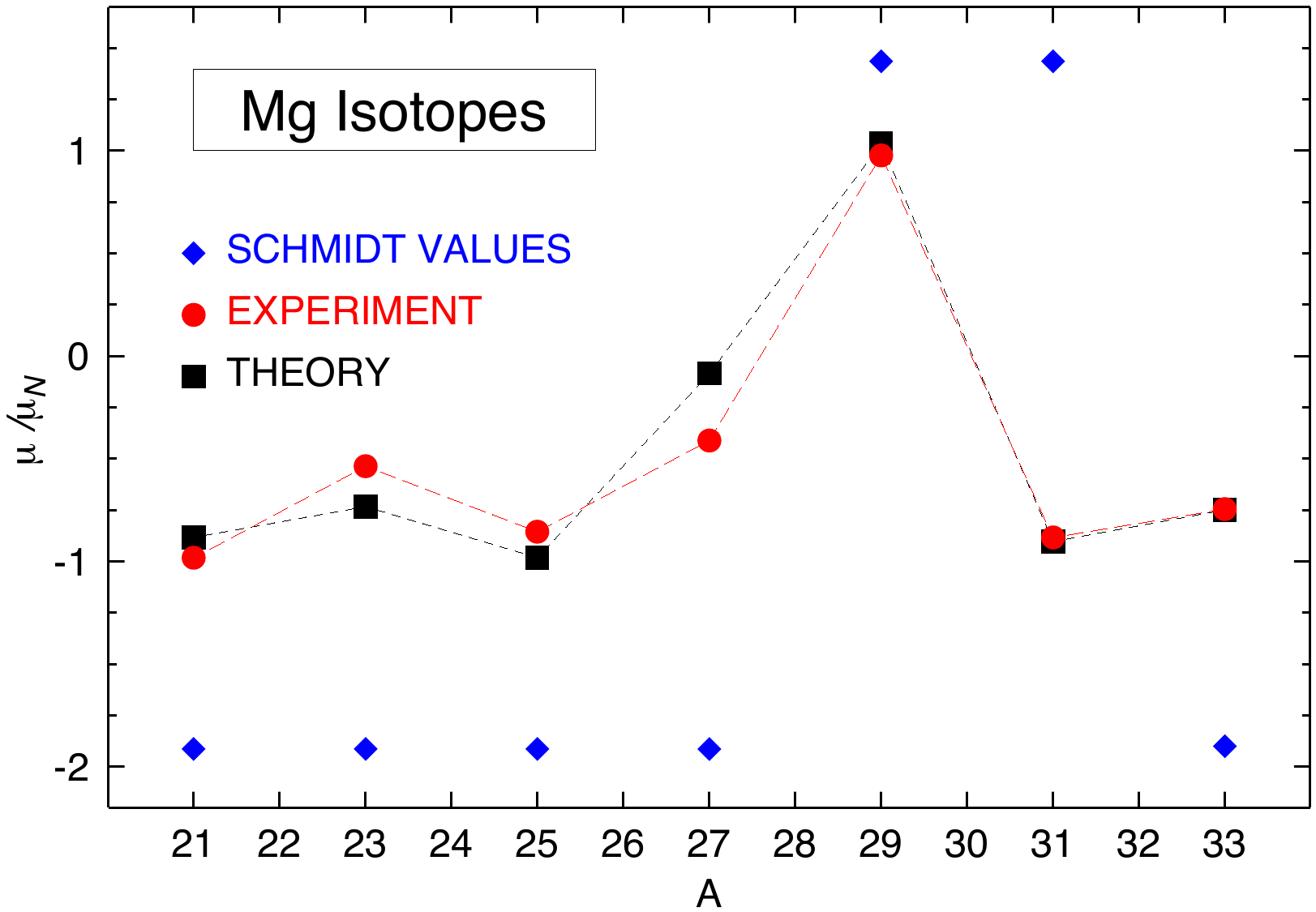}	
		\caption{Magnetic moments of the ground states of the Magnesium isotopes. The experimental results have been taken from the following references~: $^{21}$Mg \cite{mu_21Mg} $^{21}$Mg, \cite{mu_23Mg_1,mu_23Mg_2},  $^{25}$Mg\cite{mu_25Mg},  $^{27-31}$Mg \cite{mu_27-29-31Mg} and $^{33}$Mg \cite{mu_33Mg}. }
		\label{fig:mag_mom}       
	\end{center}
\end{figure}

Concerning the spectroscopic quadrupole moments of these nuclei  they have been listed in Table~\ref{Table1}.  Experimentally there are only two known values, namely, 11.4 (2) efm$^2$ in the case of $^{23}$Mg \cite{Q_23Mg} and 20.1(3) efm$^2$  for $^{25}$Mg \cite{Q_25Mg}.  Both values are somewhat smaller than the theoretical predictions 
 13.89 efm$^2$ and 22.47 efm$^2$, respectively. Concerning the magnetic moments there are more experimental data and these, together with the theoretical values, are plotted in Fig.~\ref{fig:mag_mom}.  In the calculations we have used the free gyromagnetic factors.   We have also plotted the  Schmidt values calculated with the occupations determined in the discussion of Fig.~\ref{PES_POS}. 
 As expected, due to the large deformations of these nuclei, the Schmidt values provide a poor description. For 
  $^{21-27}$Mg the  Schmidt  value is $-1.9 \mu_{N}$  and the experimental data are about half of it.  The relatively good agreement of the Schmidt with the experimental value  for $^{29}$Mg is probably
  due to the fact that this nucleus is the less deformed of all discussed isotopes. 
   According to the occupation of the last nucleon  $^{31}$Mg  should have the same  Schmidt magnetic moment  as  $^{29}$Mg. In contrast with the latter the experimental value for $^{31}$Mg, however, differs significantly from the Schmidt value.  This is probably due to the fact that $^{31}$Mg  is far more deformed ($\beta=0.60$) than $^{29}$Mg ($\beta=0.37$)  and therefore further away from the spherical limit. For  $^{33}$Mg, as for the lighter isotopes, the Schmidt value is about twice as large as the experimental data.
   Concerning our theoretical results  we observe that our values not only reproduce the tendency of the experimental data but that they are very close to them  providing in some cases quantitative agreement.

In conclusion, we have presented a novel approach with exact conservation of angular momentum and  particle number to describe odd-even nuclei. We have applied this theory to the description of ground-state properties of the Magnesium isotopic chain  with the effective Gogny force.  The results are in very good agreement with the experimental  bulk properties, energy gaps and electromagnetic moments. 

\section*{Acknowledgements}
We would like to thank Dr. Shin Watanabe for clarifying some aspects of Ref.~\cite{mass_radii}. This work was supported by the Spanish Ministerio de Econom\'ia y Competitividad under contracts FPA2011-29854-C04-04 and FPA2014-57196-C5-2-P.

%
%
%

\end{document}